\documentclass[aps,superscriptaddress,preprintnumbers,showpacs,twocolumn]{revtex4}

\usepackage{txfonts}

\usepackage{amssymb}

\usepackage{indentfirst}

\usepackage{epsfig}

\usepackage{epsf}

\usepackage{graphicx}

\def\Vec#1{{\bf #1}}

\newcommand{\be}{\begin{equation}}
\newcommand{\ee}{\end{equation}}
\newcommand{\ba}{\begin{eqnarray}}
\newcommand{\ea}{\end{eqnarray}}

\newcommand{\kp}{\mathbf{k}_\perp}

\newcommand{\p}{\perp}

\begin{document}
\newcommand*{\pku}{School of Physics, Peking University, Beijing 100871, China}\affiliation{\pku}
\newcommand*{\usm}{Departamento de F\'\i sica y Centro de Estudios
Subat\'omicos, Universidad T\'ecnica Federico Santa Mar\'\i a,
Casilla 110-V, Valpara\'\i so, Chile}\affiliation{\usm}
\newcommand*{\SKL}{State Key Laboratory of Nuclear Physics and
Technology, Peking University, Beijing 100871,
China}\affiliation{\SKL}

\title{Extracting Boer-Mulders functions from $p+D$ Drell-Yan processes }

\author{Bing Zhang}\affiliation{\pku}
\author{Zhun Lu}\affiliation{\usm}
\author{Bo-Qiang Ma}\email[Corresponding author. Electronic address: ]{mabq@phy.pku.edu.cn}\affiliation{\pku}\affiliation{\SKL}
\author{Ivan Schmidt}\email[Corresponding author. Electronic address: ]{ivan.schmidt@usm.cl}\affiliation{\usm}

\begin{abstract}\
    We extract the Boer-Mulders functions of valence and sea
quarks in the proton from unpolarized $p+D$ Drell-Yan data measured
by the FNAL E866 Collaboration. Using these Boer-Mulders functions,
we calculate the $\cos 2 \phi$ asymmetries in unpolarized $pp$
Drell-Yan processes, both for the FNAL E866/NuSea and the BNL
Relativistic Heavy Ion Collider (RHIC) experiments. We also estimate
the $\cos 2 \phi$ asymmetries in the unpolarized $p\bar{p}$
Drell-Yan processes at GSI.

\end{abstract}


\maketitle


\section{Introduction}

Recently the E866/NuSea Collaboration at FNAL has
measured~\cite{e866} the $\cos 2 \phi$ angular distribution of
Drell-Yan dimuons in $p + d$ interaction at 800 GeV/c. The magnitude
of the asymmetry turned out to be about several percent. This $\cos
2 \phi$ angular dependence, together with other angular correlations
in Drell-Yan processes, constitute remaining challenges which need
to be understood from QCD dynamics. Indeed, even before these
nucleon-nucleon interaction measurements, the NA10~\cite{na10} and
E165~\cite{conway} Collaborations had also measured the $\cos 2
\phi$ angular dependence in $\pi+N$ Drell-Yan processes. The
magnitude of the angular dependence is around 30\% for modest
transverse momenta of the lepton pair, a result which violates the
so-called Lam-Tung relation~\cite{lt78}, predicted by perturbative
QCD. Several attempts have been made to interpret these data,
including QCD vacuum effects~\cite{bnm93,bbnu} and higher-twist
mechanisms~\cite{bbkd94,ehvv94}. Furthermore, in the last decade a
significant efforts have been put forward on the understanding of the
$\cos 2 \phi$ angular dependence from the view point of the
transverse momentum dependent (TMD) Boer-Mulders function
$h_1^{\p}$~\cite{bm}. In fact, it was shown~\cite{boer} by Boer that
the angular dependence can be related to the product of two
functions $h_1^{\p}$, each of which comes from one of the incident hadrons. These $h_1^{\p}$
functions describe a correlation between the transverse spin and the
transverse momentum of a quark inside an unpolarized hadron, which
originates from inital/final state interactions~\cite{bhs02,bbh03},
and which in turn is related to the gauge invariance of the TMD
distribution functions~\cite{collins02,belitsky,bmp03}. Another
distribution having the same QCD origin of $h_1^{\p}$ is the Sivers
function~\cite{sivers}, which plays a important role in the single
spin asymmetries (SSA)~\cite{bdr} measured in semi-inclusive deep
inelastic scattering. Measurements of Sivers SSA at
HERMES~\cite{Airapetian:2004tw,hermes05} and
COMPASS~\cite{compass,compass06} have been used in order to extract
the Sivers functions by several
groups~\cite{anselmino05a,anselmino05b,efr05,cegmms,vy05}. All these
fittings show that the Sivers functions for $u$ and $d$ quarks have
similar size but opposite sign.

In contrast to the Sivers function case, detailed knowledge of
$h_1^{\p}$ is more difficult. The reason is that in processes in
which the Boer-Mulders function contributes, one finds that it is
always convoluted with itself or with other chiral-odd functions,
such as the transversity or the Collins fragmentation
function~\cite{collins93}. Despite this, some theoretical
calculations and phenomenological
analysis~\cite{gg02,bbh03,pobylista,yuan,bsy04,lm04,radici05,sissakian05,sissakian06,gamberg07,lms07,gamberg072}
have been performed on $h_1^{\p}$. An interesting problem is its
flavor dependence~\cite{lms06}. Recently lattice calculations
~\cite{lattice} and approaches based on generalized parton
distributions (GPD) ~\cite{gpds,burkardt07} suggest that the sign of
$h_1^{\p}$ for both $u$ and $d$ quarks and the corresponding magnitudes
are of similar sizes (See also the calculation in
Ref.~\cite{gamberg07}). In this paper, we will extract the
Boer-Mulders functions from the unpolarized $p+d$ Drell-Yan data
measured by the E866 collaboration, with the assumption that the
$\cos 2\phi$ angular dependence is produced solely by $h_1^\p$.
We use the simple parameterization for the Boer-Mulders function
given by
$h_1^{\p,q}(x,p_\p^2)=H_q\,x^c\,(1-x)\,f_1^q(x)\,\textrm{exp}\,(-p_\p^2/
p_{bm}^2 )$ to fit the data points of the asymmetry coefficient
$\nu$ vs $p_T$, $x_1$ and $x_2$, where q stands for the
$u,d,\bar{u},\bar{d}$ quarks. The data points of $\nu$ versus
$m_{\mu\mu}=Q^2$ and $x_F$ provide then a cross check of our fitting.
Using the resulting valence and sea quarks Boer-Mulders functions,
we estimate the $\cos 2 \phi$ asymmetries in unpolarized $pp$
Drell-Yan processes for both FNAL E866/NuSea and RHIC, and we also give
predictions for the $\cos 2 \phi$ asymmetries in unpolarized
$p\bar{p}$ Drell-Yan processes at GSI.

\section{Extracting the Boer-Mulders function from unpolarized $p+D$ Drell-Yan
data}

The angular differential cross section for unpolarized Drell-Yan
processes has the general form

\begin{eqnarray}
\frac{1}{\sigma}\frac{d\sigma}{d\Omega}&=&\frac{3}{4\pi}\frac{1}{\lambda+3}
(1+\lambda\textmd{cos}^2\theta+\mu\textmd{sin}2\theta\textmd{cos}\phi
\nonumber\\
& &
+\frac{\nu}{2}\textmd{sin}^2\theta\textmd{cos}2\phi).\label{cos2phi}
\end{eqnarray}
where $\theta$ and $\phi$ are, respectively, the polar angle and the
azimuthal angle of dileptons in the Collins--Soper frame
\cite{cs77}. The coefficients $\lambda, \mu$ and $\nu$ do not depend
on these angles, and for scattering that has azimuthal symmetry their values are $\mu
= \nu = 0$.

\begin{table}[t]
\caption{\label{tab}Best fit values of the Boer-Mulders functions}
~~ ~~~~~~~~~ ~~~~~~~~~

\begin{tabular}{|c|c|}
\hline
$H_u$           &  3.99 \\
\hline
$H_d$           &  3.83 \\
\hline
$H_{\bar{u}}$           &  0.91 \\
\hline
$H_{\bar{d}}$          &  -0.96 \\
\hline
$p_{bm}^2$          & 0.161 \\
\hline
c                   &0.45 \\
\hline
$\chi^2/d.o.f.$      & 0.79\\
\hline
\end{tabular}
\end{table}

This angular distribution has been measured in muon pair production
by pion-nucleon collisions: $\pi^-N\rightarrow\mu^+\mu^-X$, with $N$
denoting a nucleon in deuterium or tungsten, and for a $\pi^-$ beam
with energies of 140, 194, 286 GeV~\cite{na10} and 252
GeV~\cite{conway}. The experimental data show large values of $\nu$,
near 30\%.
 The most recent measurements of the angular distribution
  were performed by the E866 Collaboration~\cite{e866}, in
$p+d$ Drell-Yan processes at 800 GeV/c. The measured $\nu$ is about
several percent, a result which can not be explained by
perturbative QCD. As proposed by Boer~\cite{boer}, the non-zero
$\cos 2\phi$ term can be produced by the product of two $h_1^\p$s,
each coming from one of the two incident hadrons. In the case of the $p+D$
Drell-Yan processes, the coefficient $\nu$ can be expressed
as~\cite{lms07}
\begin{eqnarray}
\nu_{pD}=\frac{2\mathcal{F}[\,\chi \, (e_u^2 h_1^{\perp,u} +e_d^2
h_1^{\perp, d}) (h_{1}^{\perp,\bar{u}} +h_{1}^{\perp,\bar{d}})]+(q
\leftrightarrow \bar{q})}{\mathcal{F}[ (e_u^2 f_1^{u} +e_d^2 f_1^{
d}) (f_{1}^{\bar{u}} +f_{1}^{\bar{d}})]+(q \leftrightarrow
\bar{q})}\label{nud}
\end{eqnarray}
where we have used the notation
\begin{equation}
\mathcal{F}[\cdots]=\int d^2\mathbf{p}_\perp
d^2\kp\delta^2(\mathbf{p}_\perp+\kp-\mathbf{q}_\perp)\times\{\cdots\},
\end{equation}
and
\begin{equation}
\chi(\mathbf{p}_\perp,\mathbf{k}_\perp)=(2\hat{\mathbf{h}}\cdot
\mathbf{p}_\perp\hat{\mathbf{h}}\cdot \mathbf{k}_\perp
-\mathbf{p}_\perp\cdot \kp)/M^2,
\end{equation}
where $\hat{\mathbf{h}}=\Vec q_T/Q_T$ and $M$ is the mass of the
nucleon. The interacting quarks coming from the incident hadrons
have transverse momenta $\mathbf{p}_\perp$ and $\kp$, and $Q_T$
is the absolute value of the transverse momentum of the photon $\Vec q_T$.
To arrive at Eq.(\ref{nud}) we have used the isospin
relation:
\begin{eqnarray}
f^{u/D} \approx f^{u/p}+f^{u/n}=f^u+f^d.
\end{eqnarray}
for $f_1$ and $h_1^\p$. From this point of view, the angular
distribution coefficient $\nu$ measured by the E866 Collaboration
provides an opportunity to extract the Boer-Mulders function of the
nucleon. For this purpose we parameterize $h_1^\p(x,\mathbf{p}_T^2)$
in a factorized form as
\begin{equation}
h_1^{\perp,q}(x,\mathbf{p}_T^2)=h_1^{\perp,q}(x)\frac{exp(-\mathbf{p}_{T}^{2}/
p_{bm}^2)}{\pi p_{bm}^2},\label{bmexp}
\end{equation}
which is a Gaussian model for the transverse momentum dependence of
the Boer-Mulders functions, with width $ p_{bm}^2 $. As in Ref.~\cite{boer}, the
x-dependence of the Boer-Mulders functions are modeled relating its
behavior with that of the unpolarized distribution functions, as
\ba h_1^{\p,u}(x)&=&H_u\,x^c\,(1-x)\,f_1^u(x),\label{p1}\\
h_1^{\p,d}(x)&=&H_d\,x^c\,(1-x)\,f_1^d(x),\label{p2}\\
h_1^{\p,\bar{u}}(x)&=&H_{\bar{u}}\,x^c\,(1-x)\,f_1^{\bar{u}}(x),\label{p3}\\
h_1^{\p,\bar{d}}(x)&=&H_{\bar{d}}\,x^c\,(1-x)\,f_1^{\bar{d}}(x),\label{p4}
 \ea
where $f_1^q(x)$ is the well-known unpolarized integrated
distribution function. Therefore we have $H_u$,
$H_d$, $H_{\bar{u}}$,$H_{\bar{d}}$,$ p_{bm}^2 $ and $c$ as
parameters in our parametrization. The coefficient $(1-x)$ is
included in order to give the correct large-$x$
behavior~\cite{bro06} for $h_1^\perp$ compared with the unpolarized
distribution. The small-$x$ behavior of $h_1^\p$ compared with that
of $f_1(x)$ is modeled as $x^c$.

The TMD unpolarized distribution function $f_1(x,\mathbf{p}_T^2)$ is
also given in a Gaussian form
\be f_1^q(x,\mathbf{p}_T^2)=f_1^q(x)\,
\frac{\textrm{exp}\,\, (-\mathbf{p}_T^2/p_{unp}^2)}{\pi \,
p_{unp}^2 }. \label{unp} \ee

 From Eqs.~(\ref{bmexp}) to (\ref{unp}), we can de-convolute the
transverse momentum integration in Eq.~(\ref{nud}) and arrive at \ba
 &&\hspace{-15pt}\nu_{pD}(x_1,x_2,Q_T)=\frac{\
~~p_{unp}^2[\alpha]Q_T^2
\textrm{exp}\left(-\frac{Q_T^2}{2p_{bm}^2}\right )} { ~~2M^2
p_{bm}^2[\beta]\textrm{exp} \left (-\frac{Q_T^2}{2p_{unp}^2}\right
)}, \label{pdptbm}\ea
 where
 \ba
&&\hspace{-30pt}[\alpha]=x_1^c(1-x_1)x_2^c(1-x_2)(4H_uf_1^u(x_1)+H_df_1^d(x_1))\nonumber\\
&&(H_{\bar{u}}f_1^{\bar{u}}(x_2)+H_{\bar{d}}f_1^{\bar{d}}(x_2))
 +(q
\leftrightarrow \bar{q}),\nonumber \\
\label{together}\\
&&\hspace{-30pt}[\beta]=(4f_1^u(x_1)+f_1^d(x_1))(f_1^{\bar{u}}(x_2)+f_1^{\bar{d}}(x_2))\nonumber\\
&&+(q \leftrightarrow \bar{q}) \ea

Therefore the coefficient $\nu$ vs $Q_T$ can be obtained from
(\ref{pdptbm}) by integrating the numerator and the denominator over
$x_1$ and $x_2$ over, respectively: \ba
 \nu_{pD}(Q_T)=\frac{ p_{unp}^2  \int dx_1 \int dx_2
[\alpha] Q_T^2 \textrm{exp}\left(-\frac{Q_T^2}{2  p_{bm}^2 }\right
)} {2M^2  p_{bm}^2  \int dx_1 \int dx_2 [\beta]\textrm{exp} \left
(-\frac{Q_T^2}{2p_{unp}^2}\right )}, \ea In this equation, $x_1$,
$x_2$ and $Q_T$ satisfy the relation $x_1x_2s=Q^2+Q_T^2$, where $s$
and $Q$ are the c.m energy squared and the invariant mass of the
lepton pair respectively. In a similar way we can obtain expressions
for $\nu_{pD}(x_1)$ and $\nu_{pD}(x_2)$.

The E866/NuSea Collaboration has measured $\nu_{pd}$ vs $Q_T$,
$x_1$, $x_2$, $x_F$ and $m_{\mu\mu}$ in the following kinematical region:
\begin{eqnarray}
&&4.5\,\textrm{GeV}<Q
<9\,\textrm{GeV}~~\textrm{and}~~10.7\,\textrm{GeV}<Q<15\,\textrm{GeV},\nonumber\\
&&0.1< x_1 <0.9,~~~~0.02<x_2<0.24 \nonumber\label{e866cut}
\end{eqnarray}
We therefore use the above expressions for $\nu_{pd}(Q_T)$,
$\nu_{pd}(x_1)$ and $\nu_{pd}(x_2)$ in order to fit the experimental data of
$\nu$ vs $Q_T$, $x_1$ and $x_2$ in $pD$ processes measured by
E866/NuSea Collaboration. We perform the fitting using the MINUIT program. For
the unpolarized distribution function $f_1^q(x)$ we adopt the
MRST2001 (LO set) parametrization ~\cite{mrst}, with QCD evolution
taken into account. We use $ p_{unp}^2 = 0.25$, following the
choice in Ref.~\cite{punp}, a value which was obtained by fitting the
azimuthal dependence of the SIDIS unpolarized cross section. Notice
that these values are assumed to be constant and flavor independent.

\begin{figure*}
\begin{center}
\scalebox{0.33}{\includegraphics[0pt,70pt][300pt,220pt]{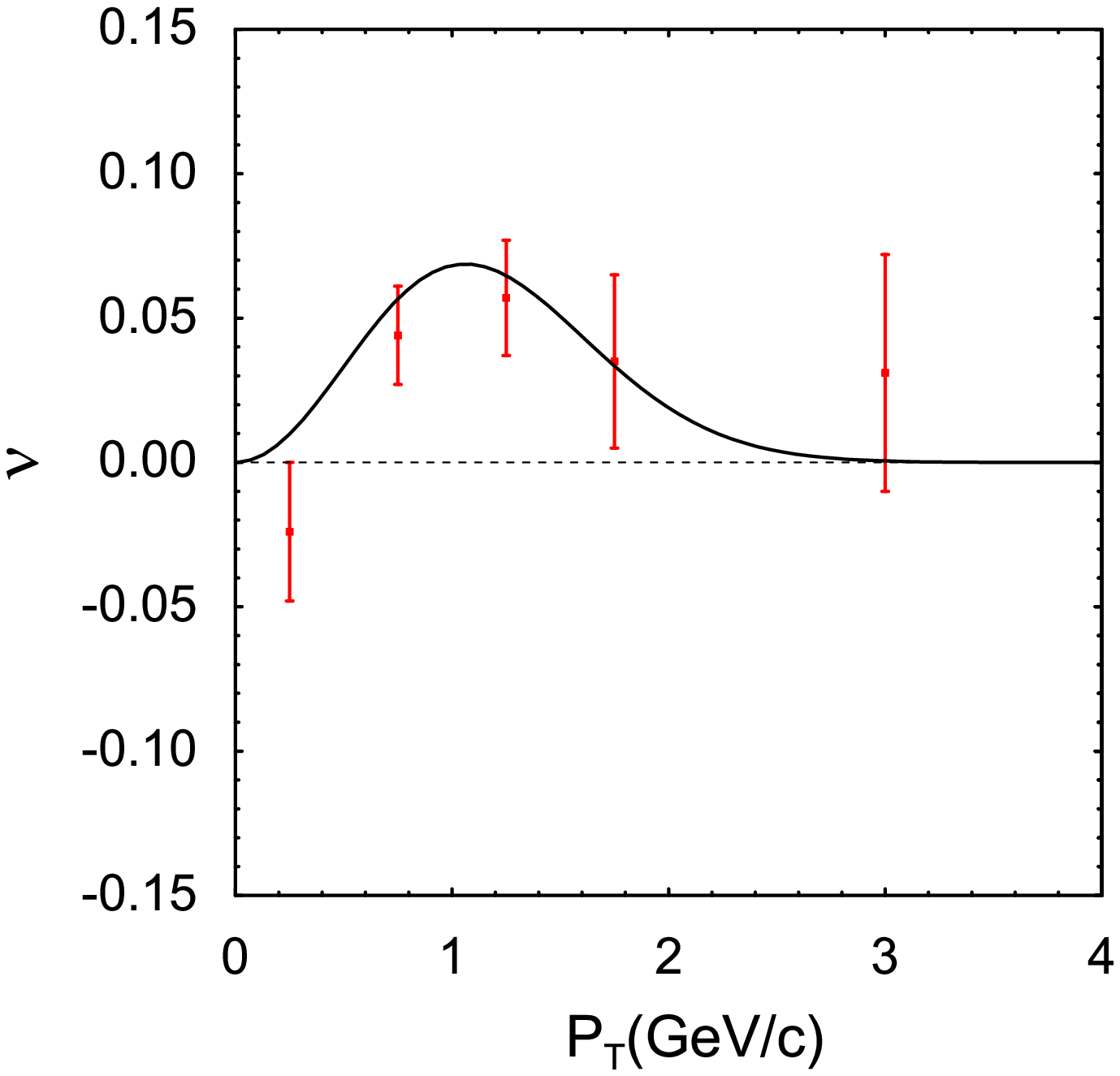}}
\scalebox{0.33}{\includegraphics[-120pt,70pt][490pt,460pt]{pdx1.eps}}
\scalebox{0.33}{\includegraphics[80pt,70pt][490pt,460pt]{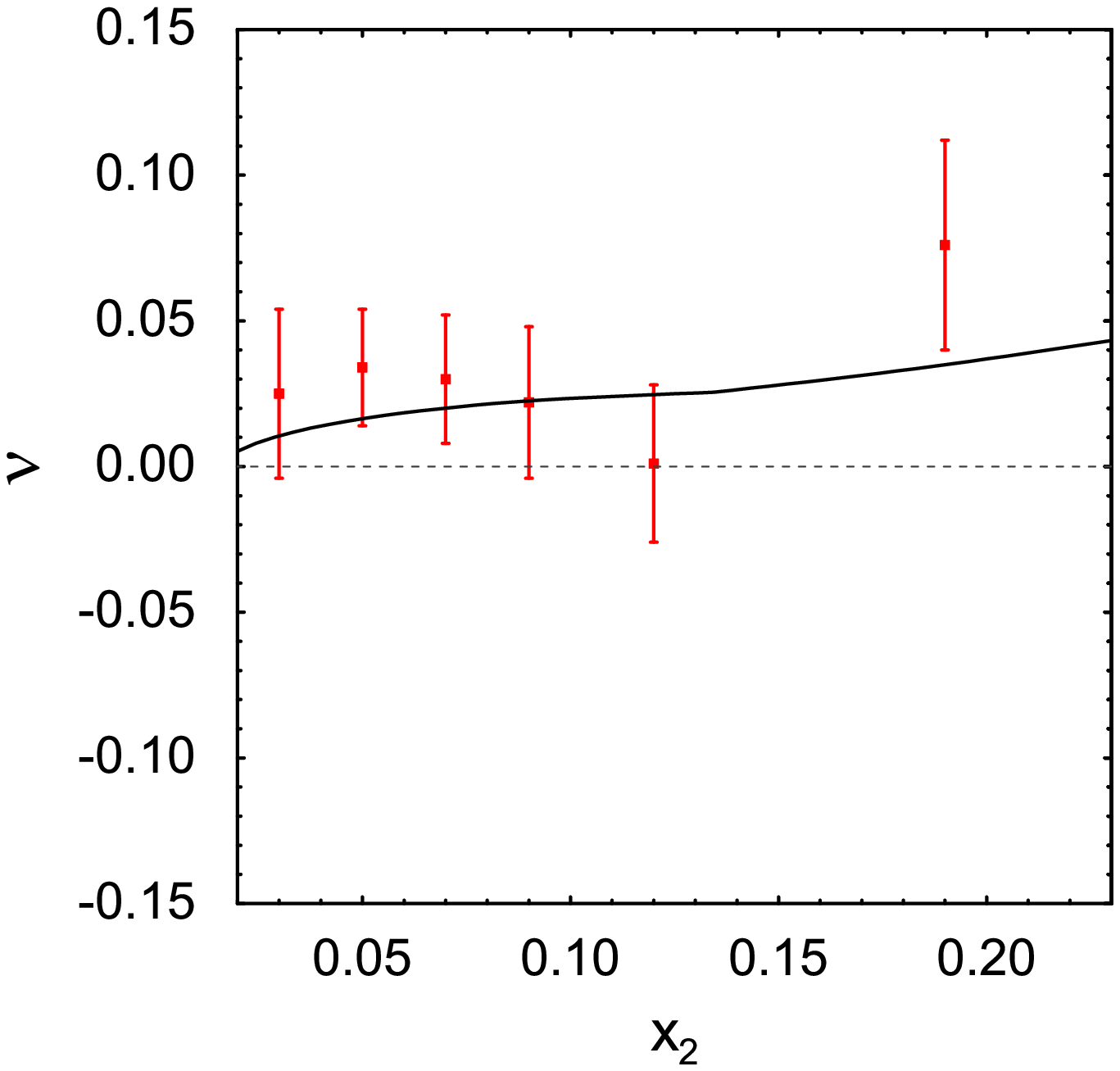}}
\caption{\small Fits to the $p_T$,$x_1$, $x_2$-dependent $\cos 2
\phi$ asymmetries $\nu_{pD}$ for Drell-Yan processes. Data are from
the FNAL E866/NuSea collaboration.}\label{pdptx1x2}
\end{center}
\end{figure*}

\begin{figure}
\begin{center}
\scalebox{0.9}{\includegraphics{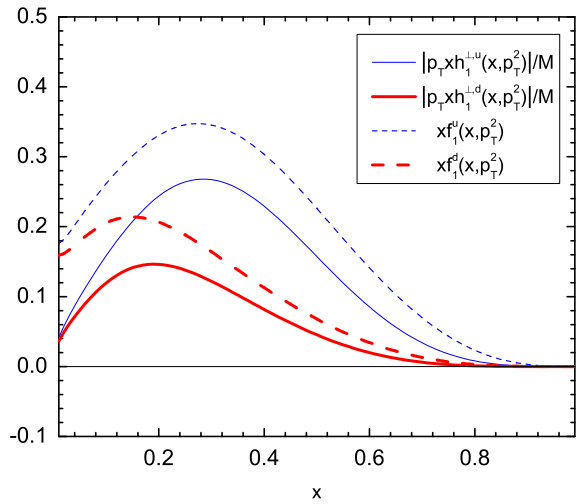}}
 \caption{\small Comparison of $|p_{T} x  h^\p_1(x,\mathbf{p}
 _{T}^2)|/M$ and $ x f_1(x,\mathbf{p}_{T}^2)$ for $u$ and $d$ quarks
 at $p_{T}=0.45$ GeV and $Q= 1$ GeV.
 Here $f_1$ is a combination of valence and sea quark
 distributions.
 }\label{soffersbound}
\end{center}
\end{figure}

 The best fitting values for the parameters are given in
 Table~\ref{tab}, and in Fig.~\ref{pdptx1x2} we show the fitting result compared with the
E866/NuSea data. Notice that the results for the Boer-Mulders
functions given in Table~\ref{tab} are obtained for the Drell-Yan
process, and to obtain the corresponding Boer-Mulders functions in
SIDIS one should reverse their signs~\cite{collins02}. Therefore our
results agree with the expectation that the Boer-Mulders functions
for u and d quarks in SIDIS are negative and have the same sign. We
also want to emphasize that the Boer-Mulders functions we extracted
are within a positive bound~\cite{soffers} \be \frac{|p_{T}
h^\p_1(x,\mathbf{p}
 _{T}^2)|}{M} \le f_1(x,\mathbf{p}_{T}^2).
 \ee
 We also find that at
 \be p_{T}=\sqrt{p_{unp}^2 \,
 p_{bm}^2/(2(p_{unp}^2-p_{bm}^2))}=0.45 \textrm{GeV},
 \ee
 the ratio $p_{T} h^\p_1(x,\mathbf{p}
 _{T}^2)/(Mf_1(x,\mathbf{p}_{T}^2))$ has its maximum for all $x$.
 As an example, in
 Fig.~\ref{soffersbound} we show a comparison of $p_{T} x h^\p_1(x,\mathbf{p}
 _{T}^2)/M$ and $x f_1(x,\mathbf{p}_{T}^2)$, for both $u$ and $d$ quarks at $p_{T}=0.45$
 GeV and $Q= 1$ GeV, and therefore we conclude that our results obey the positive bound for all
 $p_T$.

\begin{figure}
\begin{center}
\scalebox{0.3}{\includegraphics[90pt,70pt][490pt,460pt]{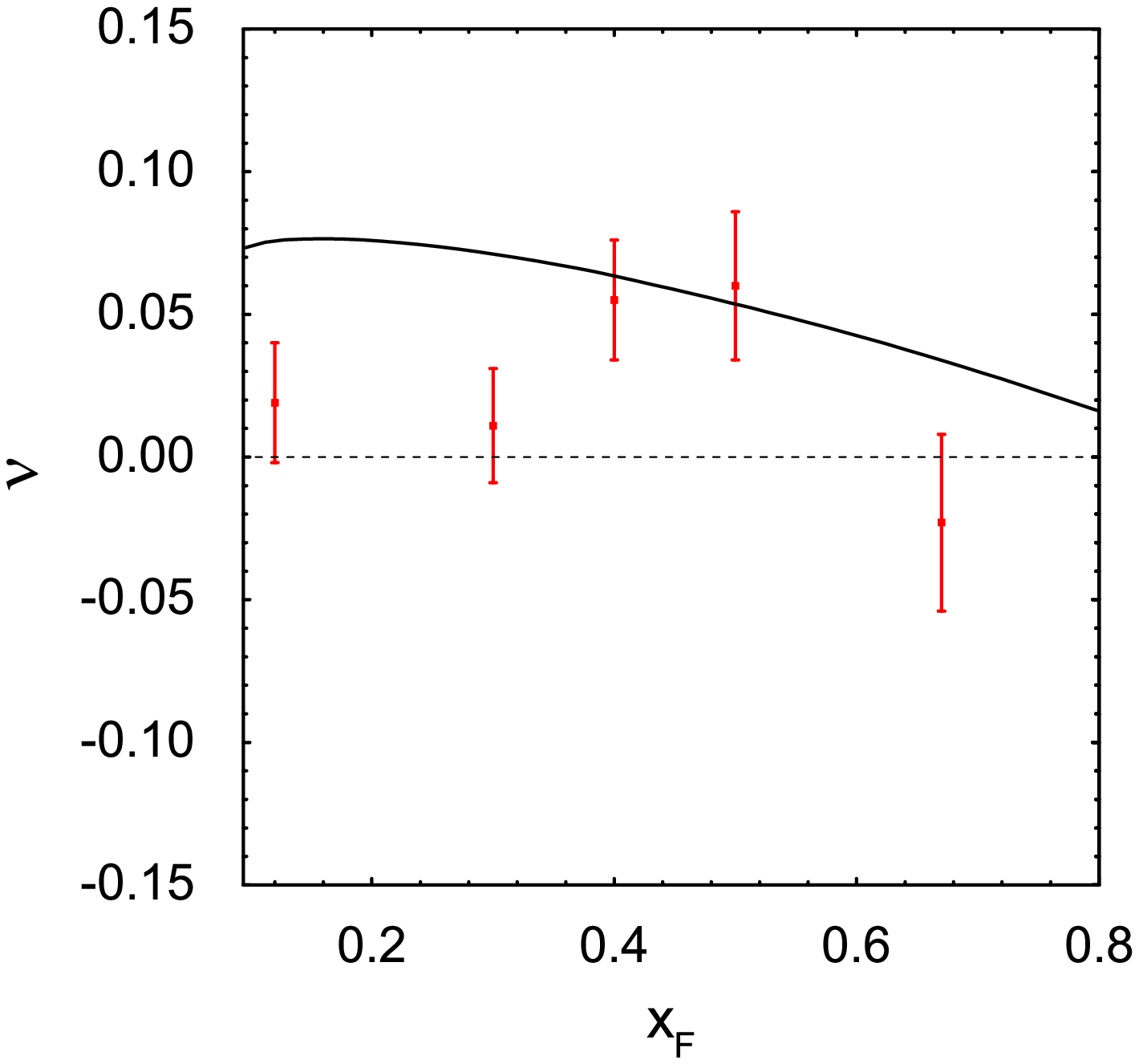}}
\scalebox{0.3}{\includegraphics[90pt,70pt][490pt,460pt]{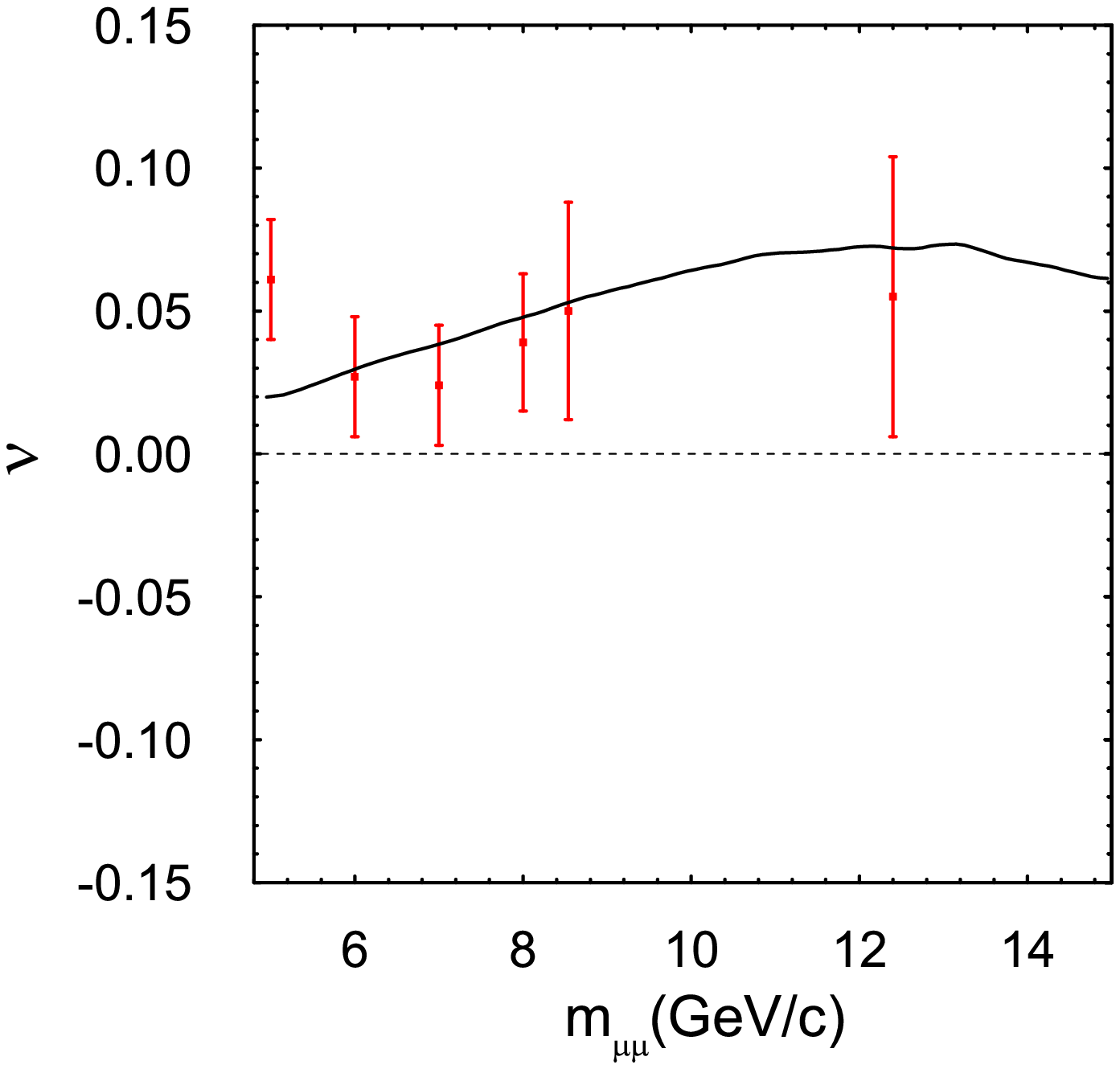}}
 \caption{\small The
$Q$-dependent $\cos 2 \phi$ asymmetry $\nu_{pD}$ for
Drell-Yan processes at FNAL E866/NuSea, presenting both the experimental data
and the results we estimate with the best fit values of
the Boer-Mulders functions in Table~\ref{tab} (line).}\label{mmumu}
\end{center}
\end{figure}

As a cross check, we give results for the coefficient $\nu$ versus $x_F$ and $m_{\mu\mu}$
in $pD$ Drell-Yan processes for E866/NuSea, using the extracted
Boer-Mulders functions and comparing them with data, as shown in
Fig.~\ref{mmumu}. In the calculation we use the relation $x_{1/2} =
(\pm x_F + (Q^2+Q_T^2)/s)/2$. The results in the figure show good
agreement with data.

\begin{figure}
\begin{center}
\scalebox{0.4}{\includegraphics[280pt,30pt][490pt,460pt]{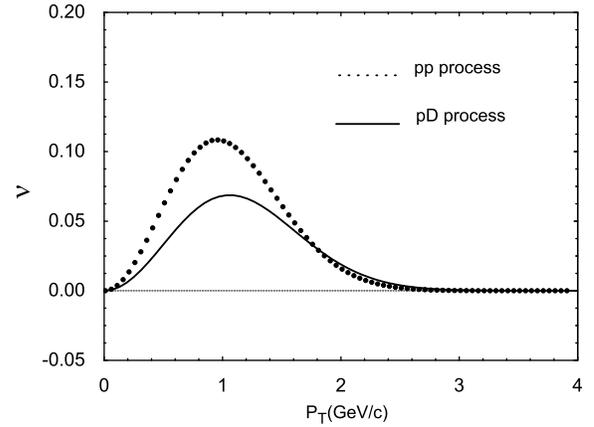}}
\caption{\small The $p_T$-dependent $\cos 2 \phi$ asymmetries $\nu$
in both $pp$ (dotted curve)and $pD$(solid curve) Drell-Yan processes
at FNAL E866/NuSea, calculated with the fitted Boer-Mulders
functions presented in Table~\ref{tab}.}\label{pppt}
\end{center}
\end{figure}

\section{Predictions for $pp$ and $p\bar{p}$ process}

In the previous section we have performed a fitting on the
unpolarized $pd$ Drell-Yan data, and obtained a set of Boer-Mulders
functions. In this section, we will use the extracted functions to
predict the result in other processes. We first focus on unpolarized
$pp$ Drell-Yan processes. The $\cos 2 \phi$ angular distribution in
this process can be also measured at E866/FNAL, and the expression
that we get for $\nu$ using our parametrization of the Boer-Mulders
functions is

\ba
 &&\hspace{-15pt}\nu_{pp}(x_1,x_2,Q_T)=\frac{\
~~p_{unp}^2[\kappa]Q_T^2
\textrm{exp}\left(-\frac{Q_T^2}{2p_{bm}^2}\right )} { ~~2M^2
p_{bm}^2[\lambda]\textrm{exp} \left (-\frac{Q_T^2}{2p_{unp}^2}\right
)}, \label{ppptbm}\ea
 where
 \ba
&&\hspace{-30pt}[\kappa]=x_1^c(1-x_1)x_2^c(1-x_2)(4H_uf_1^u(x_1)H_{\bar{u}}f_1^{\bar{u}}(x_2)\nonumber\\
&&+H_df_1^d(x_1)H_{\bar{d}}f_1^{\bar{d}}(x_2))+(q\leftrightarrow \bar{q}),\nonumber \\
&&\hspace{-30pt}[\lambda]=(4f_1^u(x_1)f_1^{\bar{u}}(x_2)+f_1^d(x_1)f_1^{\bar{d}}(x_2))\nonumber\\
&&+(q \leftrightarrow \bar{q}) \ea

With the Boer-Mulders functions for both valence and sea quarks in the
proton that we obtained above, and taking the same integration regions
for $Q,x_1,x_2$ in evaluating Eq. ~(\ref{ppptbm}), we get the results for
$\cos 2\phi$ asymmetry in unpolarized $pp$ and $pD$ Drell-Yan
processes that are shown in Fig.~\ref{pppt} (solid and dashed lines
respectively).

The results in these two processes present little difference between
them.

\begin{figure}
\begin{center}
\scalebox{0.36}{\includegraphics[280pt,70pt][490pt,460pt]{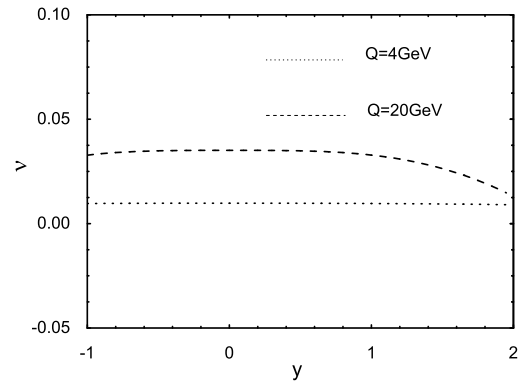}}
\caption{\small The $y$-dependent $\cos 2 \phi$ asymmetries $\nu$
for $pp$ Drell-Yan process at RHIC, with the kinematical conditions
$\sqrt{s}=200$ GeV, $-1< y <2$, and calculated with the Boer-Mulders
functions that we got. The dotted and dashed lines show the
asymmetries at $Q=4$ and 20 GeV, respectively.}\label{ppyrhic}
\end{center}
\end{figure}

Another promising testing ground of the $\cos 2 \phi$ asymmetry in
unpolarized $pN$ Drell-Yan processes is BNL RHIC~\cite{rhic},
since this experiment is also feasible at this accelerator.
With the Boer-Mulders functions given in Table~\ref{tab}, we
also estimate the $\cos 2 \phi$ asymmetry $\nu$ at RHIC for an
unpolarized experiment, with kinematical constraints $\sqrt{s}=200$
GeV and $-1 < y <2$. Here $y$ is the rapidity defined as
$y=\frac{1}{2}\textrm{ln}\frac{x_1}{x_2}$, with
$x_{1/2}=\sqrt{\frac{Q^2+Q_T^2}{s}}\,e^{\pm y}$. After integrating
over $x_1$ in Eq.(\ref{ppptbm}), we get the $Q_T$-dependent asymmetries
for $Q=4$ GeV (solid line) and $Q=20$ GeV (dashed line)
shown in Fig.~\ref{ppyrhic}.

\begin{figure}
\begin{center}
\scalebox{0.75}{\includegraphics[30pt,15pt][550pt,400pt]{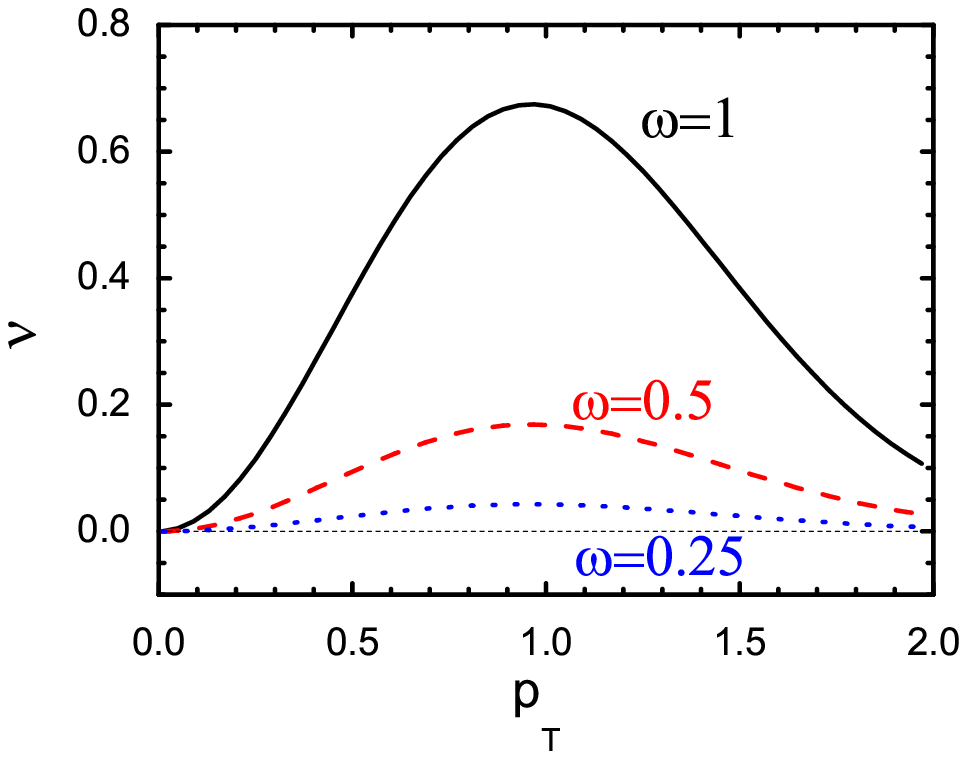}}
\caption{\small The $p_T$-dependent $\cos 2 \phi$ asymmetries $\nu$
for $p\bar{p}$ Drell-Yan process at GSI, with the kinematics:  c.m.
energy $s=45~\textrm{GeV}^2$, and invariance mass square of the
lepton pair $Q^2=2.5~ \textrm{GeV}^2$. The solid, dashed and dotted
curves correspond to the free coefficient $\omega$=1, 0.5, 0.25
respectively.}\label{ppbargsi}
\end{center}
\end{figure}
We should notice that in the extraction of the Boer-Mulders
functions with the MINUIT program, the coefficients of Boer-Mulders functions for
valence quarks $H_u$ and $H_d$ are always coupled with the
coefficients of sea quarks  $H_{\bar{u}}$ and $H_{\bar{d}}$, as seen in
Eq.~(\ref{together}). We cannot separate them except
by introducing free coefficients $\omega_q$ with possible flavor
dependence.

With the free coefficient $\omega_u=\omega_d=\omega$, the
Boer-Mulders functions are modeled comparing its behavior with that
of the unpolarized distribution functions, as
\ba h_1^{\p,u}(x)&=&\omega H_u\,x^c\,(1-x)\,f_1^u(x),\label{p1new}\\
h_1^{\p,d}(x)&=&\omega H_d\,x^c\,(1-x)\,f_1^d(x),\label{p2new}\\
h_1^{\p,\bar{u}}(x)&=&\frac{1}{\omega}H_{\bar{u}}\,x^c\,(1-x)\,f_1^{\bar{u}}(x),\label{p3new}\\
h_1^{\p,\bar{d}}(x)&=&\frac{1}{\omega}H_{\bar{d}}\,x^c\,(1-x)\,f_1^{\bar{d}}(x),\label{p4new}
 \ea

   The results we get above for the $\nu_{pD},\nu_{pp}$ are independent
of the free coefficient $\omega$ because when we use
Eqs.~(\ref{p1new})-(\ref{p4new}) into the $\nu_{pD}$, $\nu_{pp}$ in
Eqs.~(\ref{pdptbm}) and (\ref{ppptbm}), the free coefficient $\omega$
cancels in the product of sea and valence components.

   Recently the $\cos 2\phi$ azimuthal asymmetry of $p\bar{p}$
Drell-Yan processes has also received much attention, since there have been
proposals to study spin phenomena in polarized and unpolarized
$p\bar{p}$ scattering at the High-Energy Storage Ring (HESR) of GSI
\cite{PAX,PANDA}. The $p_T$, $x_1$ and $x_2$-dependent $\cos 2\phi$
asymmetry $\nu_{p \bar{p}}$, obtained using our parametrization
for the Boer-Mulders functions with the free coefficient $\omega$ are
\ba
 &&\hspace{-15pt}\nu_{p\bar{p}}(x_1,x_2,Q_T)=\frac{\
~~p_{unp}^2[\varsigma]Q_T^2
\textrm{exp}\left(-\frac{Q_T^2}{2p_{bm}^2}\right )} { ~~2M^2
p_{bm}^2[\tau]\textrm{exp} \left (-\frac{Q_T^2}{2p_{unp}^2}\right
)}, \ea
 where
 \ba
&&\hspace{-30pt}[\varsigma]=\omega^2x_1^c(1-x_1)x_2^c(1-x_2)(4H_uf_1^u(x_1)H_uf_1^u(x_2)\nonumber\\
&&+H_df_1^d(x_1)H_df_1^d(x_2))+(q\leftrightarrow \bar{q}),\nonumber \\
&&\hspace{-30pt}[\tau]=(4f_1^u(x_1)f_1^u(x_2)+f_1^d(x_1)f_1^d(x_2))+(q\leftrightarrow
\bar{q}) \label{ppbarbm} \ea

In the expression above, we use charge-conjugation relations such
as:
\begin{eqnarray}
&&f_1^{\bar{u}/\bar{p}}=f_1^{u/p},~~~f_1^{\bar{d}/\bar{p}}=f_1^{d/p}\nonumber\\
&&f_1^{u/\bar{p}}=f_1^{\bar{u}/p},~~~f_1^{d/\bar{p}}=f_1^{\bar{u}/p}.
\end{eqnarray}

  In Eq.(\ref{ppbarbm}), we can see that the result for this $p\bar{p}$ process
depends on the free coefficient $\omega$. Therefore we give predictions
for $\nu_{p \bar{p}}$ at GSI, as shown in Fig.~\ref{ppbargsi}, for
three different values of $\omega$. The kinematics in GSI can be
chosen as c.m. energy $s=45~\textrm{GeV}^2$ and $Q^2=2.5~
\textrm{GeV}^2$, where the $x$ is in the valence region and the
asymmetry can be larger than the asymmetry in $pp$ processes at
E866/NuSea and RHIC. The coming experimental data can
fix $\omega$ to give the exact Boer-Mulders functions for both quarks
and antiquarks.

\section{Conclusion}

In summary, we have extracted the Boer-Mulders functions of valence
and sea quarks inside the proton from unpolarized $p+D$ Drell-Yan
data at 800 Gev/c, measured by the FNAL E866/NuSea Collaboration.
With the Boer-Mulders functions that we get, we estimated the $\cos
2\phi$ asymmetries in unpolarized $pp$ Drell-Yan processes at
E866/NuSea and RHIC.  We also presented an estimation of the $\cos
2\phi$ azimuthal asymmetry in $p\bar{p}$ Drell-Yan processes at GSI.
We hope that future measurement for the $\cos 2 \phi$ asymmetry at
$p p$ and $p\bar{p}$, as well as in SIDIS, can help to pin down
the Boer-Mulders functions for the valence and sea quarks.

{\bf Acknowledgements.} We acknowledge Jen-Chieh Peng and Lingyan
Zhu for providing the E866/NuSea data points and for discussions.
This work is partially supported by National Natural Science
Foundation of China (Nos.~10721063, 10575003, 10505011, 10528510),
by the Key Grant Project of Chinese Ministry of Education
(No.~305001), by the Research Fund for the Doctoral Program of
Higher Education (China).

\end{document}